\documentclass[journal,transmag]{IEEEtran}
\usepackage{graphicx}
\usepackage{cite}
\usepackage[ruled,vlined,linesnumbered]{algorithm2e}
\usepackage{picinpar}
\usepackage{amsmath}
\usepackage{amssymb}
\usepackage{stfloats}
\usepackage{url}
\usepackage[latin1]{inputenc}
\usepackage{colortbl}
\usepackage{soul}
\usepackage{multirow}
\usepackage{pifont}
\usepackage{color}
\usepackage{alltt}
\usepackage[hidelinks]{hyperref}
\usepackage{enumerate}
\usepackage{siunitx}
\usepackage{breakurl}
\usepackage{epstopdf}
\usepackage{pbox}
\usepackage{float}
\usepackage{subcaption}

\begin{document}
\title{An overview of sensing platform-technological aspects for vector magnetic measurement: a case study of the application in different scenarios}

\author{\IEEEauthorblockN{Huan Liu\IEEEauthorrefmark{1,2,3,4},~Haobin Dong\IEEEauthorrefmark{1,2,3}, Jian Ge\IEEEauthorrefmark{1,2,3}, and Zheng Liu\IEEEauthorrefmark{4}}
\vspace{0.5cm}
\IEEEauthorblockA{\IEEEauthorrefmark{1}School of Automation,
China University of Geosciences, Wuhan, 430074, China}
\IEEEauthorblockA{\IEEEauthorrefmark{2}Hubei Key Laboratory of Advanced Control and Intelligent Automation for Complex Systems, Wuhan, 430074, China}
\IEEEauthorblockA{\IEEEauthorrefmark{3}Engineering Research Center of Intelligent Technology for Geo-Exploration, Ministry of Education, Wuhan, 430074, China}
\IEEEauthorblockA{\IEEEauthorrefmark{4}School of Engineering, University of British Columbia Okanagan, Kelowna BC, V1V 1V7, Canada}
\thanks{Corresponding author: Haobin Dong (email: donghb@cug.edu.cn).}}


\IEEEtitleabstractindextext{
\begin{abstract}
Magnetic sensing platform techniques have been used in many years in an attempt to better evaluate the likelihood of recoverable hydrocarbon reservoirs by determining the depth and pattern of sedimentary rock formations containing magnetic minerals, such as magnetite. Utilizing airplanes, large-area magnetic surveys have been conducted to estimate, for example, the depth of igneous rock and the thickness of sedimentary rock formations. In this case, the vector magnetic survey method can simultaneously obtain the modulus and direction information of the Earth's magnetic field, which can effectively reduce the multiplicity on data inversion, contribute to the quantitative interpretation of the magnetic body and obtain more precise information and characteristics of magnetic field resource, so as to improve the detection resolution and positioning accuracy of the underground target body. This paper presents a state-of-the-art review of the application situations, the technical features, and the development of the vector magnetic sensing platform-technical aspects for different application scenarios, i.e., ground, wells, marine, airborne, and satellites, respectively. The potential of multi-survey sensing platform technique fusion for magnetic field detection is also discussed. 
\end{abstract}

}

\markboth{Measurement}{}

\maketitle
\IEEEdisplaynontitleabstractindextext
\IEEEpeerreviewmaketitle

\section{Introduction}
Magnetic sensing platform techniques are one of the most effective methods for geophysical engineering and environmental exploration~\cite{Maus-2009-J,Liu-2020-J1,Denisov-2014-J,Liu-2020-J2}, e.g., unexploded ordnance (UXO) detection~\cite{Chen-2020-J,Chen-2019-J}, mineral exploration~\cite{Liu-2019-J2,Tan-2019-J,Liu-2020-J}, etc~\cite{Dong-2017-J,Winslow-2012-J,Liao-2019-J,Zeng-2018-J}. In recent years, the vector magnetic survey techniques have developed rapidly. When compared it with the traditional scalar magnetic survey methods~\cite{Liu-2017-J2, Liu-2018-J2, Liu-2018-J4}, the vector magnetic survey can directly provide vector magnetic field information and reveal detailed features of the Earth's magnetic field without computing from the scalar measurements, which can effectively reduce the multiplicity on data inversion~\cite{Hoffman-1983-J, Vavassori-2000-J, Mohamadabadi-2014-J}. 

For now, the Overhauser magnetometer and optically pumped magnetometer are the two main scalar instruments. Without the deficiencies of a dead zone and heading errors that affect optically pumped sensors, the Overhauser sensor has wider application prospects. Although scalar measurements are characterized by excellent performance, e.g., the resolution and the accuracy of a commercial Overhauser magnetometer are 0.01 nT and 0.1 nT, respectively, the magnetic information obtained is limited. The vector magnetometers commonly used in geophysical engineering mainly include fluxgate magnetometers and superconducting magnetometers~\cite{Canciani-2017-J,Stele-2020-J}. The fluxgate magnetometer is one of the most used vector instruments, with a resolution around 1 nT, which is worse than that of the scalar instruments. 

Since the vector magnetic survey data includes the magnetic field value and the corresponding direction, the data obtained from the magnetic survey always needs to be corrected to the geographic coordinate system, corrected the drift with temperature, etc., for further processing and interpretation. For instance, in the airborne magnetic survey, it is necessary to obtain the system attitude and orientation during flight, i.e., an error of 10$^{-3}$$^{\circ}$ of one of the inertial measurement unit angle could produce an error of 1 nT for components after rotation form magnetometer coordinate system to the Earth coordinate system. Hence, the final magnetic measurement accuracy always depends on the accuracies of the magnetometer and the attitude measurement system, respectively~\cite{Dou-2016-J, Calou-2020-J}. 

In literature~\cite{Blakely-1996-J}, the total field anomaly can be transformed into some other components of the magnetic field. However, such transformations will be limited for certain ambient-field directions in the same way that reduction to the pole is limited as low latitudes. Consider, for example, a total-field anomaly measured near the magnetic equator and caused by a body with purely induced magnetization. Both unit vectors in the direction of the magnetization and in the direction of the ambient field will have shallow inclinations in this case, and transforming the total-field anomaly into the vertical component of the magnetic field can be expected to be an unstable operation. Further, any noise within the measurements will generate artifacts, typically short in wavelength and elongated parallel to the declination. Hence, the vector magnetic survey would be a better choice in this case because it can obtain the vector magnetic field data directly.

To further suppress the uncontrollable influence, such as the background noise, the geomagnetic diurnal change, etc., the gradiometer based magnetic vector survey technique can be employed to measure the difference between readings from dual magnetometers. Using gradiometers we can simply cancel out the spatially homogeneous part of the background field vector. Furthermore, the magnetic gradient is much less susceptible to orientation error compared to the magnetic vector. To sum up, the vector magnetic survey, especially the corresponding gradiometer based approach, can contribute to the quantitative interpretation of the magnetic body and obtain more precise information and characteristics of magnetic field resource, and thus improve the detection resolution and positioning accuracy of the underground target body~\cite{Hurwitz-1960-J, Merayo-2000-J, Jeng-2017-J}. 

The vector magnetic survey technique can be mainly divided into ground magnetic survey, wells magnetic survey, marine magnetic survey, airborne magnetic survey, and satellites magnetic survey, and each technique has its own characteristics. The ground magnetic survey is mainly used for the detection of ore bodies distributed horizontally on the surface~\cite{Liu-2020-J3}. However, due to the abnormal superposition of different geological bodies on the surface, it is difficult to distinguish the depth. Hence, it is often combined with other magnetic survey techniques~\cite{Liu-2018-J1}. The wells magnetic survey is based on the magnetic characteristics of rock ore, measuring three orthogonal components of the geomagnetic field, and the radial detection range is large~\cite{Liu-2013-J1}. Further, this technique can be employed to find both strong magnetite deposits and non-ferrous metals with weaker magnetic intensities, and it is an effective method for detecting magnetic ore bodies, especially, provides a scientific basis for the exploration and evaluation of mineral resources in the deep earth~\cite{Vouillamoz-2002-J}. The marine magnetic survey always adopts a ship equipped with magnetometers to conduct geomagnetic surveys in the ocean area. It plays a crucial role in the detection of military targets such as underwater submarines, unexploded weapons, and magnetic obstacles~\cite{Fan-2017-J}. The traditional approach is mainly based on total field measurement. 

In recent years, the three-component or full-tensor magnetic gradient based marine magnetic survey technique has been employed to obtain more geomagnetic information to provide important parameters for the naval battlefield~\cite{Ge-2020-J}. The airborne magnetic survey is appropriate for scanning large areas during reconnaissance to delimit target areas for detailed ground surveys during the prospecting stage~\cite{Doll-2003-J}. In particular, it can be used in coalfield studies to map out a broad structural framework over an exploration area with complex terrain both quickly and effectively~\cite{Barros-2016-J}. The satellite magnetic survey can obtain high-quality, global coverage magnetic survey data, conduct all-weather and uninterrupted measurements, and thus carry out a series of scientific researches such as the evolution of the geomagnetic field and the law of space current movement~\cite{Mayhew-1980-J,Ia-1997-J}.

In this paper, according to different application scenarios, first, the technical characteristics of the vector magnetic measurement compared with the scalar magnetic measurement are summarized. Second, we elaborate on the application situations, the technical features, and the instruments development of the ground vector magnetic survey technique, the wells vector magnetic survey technique, the marine vector magnetic survey technique, the airborne vector magnetic survey technique, and the satellites vector magnetic survey technique, respectively. Finally, the advantages and disadvantages of these five vector magnetic survey techniques are summarized, and their application prospects and development directions are discussed. 

\section{Application in Different Scenarios}
\subsection{Ground vector magnetic measurement}
The magnetic field at any point in space is a vector quantity, which means there is a direction associated with the field as well as a field strength. The direction of the arrow can be thought of as the direction of the magnetic field. The length of the arrow can be thought of as the strength of the field, i.e. the longer the arrow, the stronger the field. We can define this length as $B$, which implies that $B$ represents the strength of the magnetic field. Generally, a single axis measuring device will change its reading depending on which way the sensitive axis is oriented with respect to the direction of the magnetic field. To obtain a complete representation of magnetic field at any point in space, one needs not only the value of $B$, but the direction, which can be expressed as the three components, $B_x, B_y$ and $B_z$. Some magnetic field sensors measure only one component of the magnetic field, e.g., Fluxgates and Hall effect instruments which are referred to as single axis devices~\cite{Ripka-1992-J,Ripka-2000-J}. Other instruments measure only the total field $B$, e.g., NMR~\cite{Liu-2017-J4,Liu-2018-J3}, ESR~\cite{Overhauser-1953-J,Ge-2016-J}. Aiming to get more information for engineering geophysical exploration, it is possible to combine single-axis sensors to give three field measurements in a single probe package. These are referred to as three-axis devices~\cite{Ripka-2003-J,Ripka-2010-J}. 

The ground vector magnetic survey using three-axis device is mainly divided into two types of measurement: 1) stationary measurement and 2) mobile measurement. The purpose is to analyze the geomagnetic characteristics by measuring the Earth's magnetic field. The main task is to survey ore bodies with magnetic anomalies, geological structural zoning, and geological mapping services~\cite{Liu-2016-J1,Liu-2016-J2,Liu-2019-J5,Kvamme-2003-J,Goldberg-2017-J}. 

The stationary measurement has been widely used in many fields such as the basic network of seismic stations, underground magnetic anomaly source detection, and weak magnetic resonance signal acquisition, because of its early development and easy implementation~\cite{Liu-2019-J3,Liu-2019-J4}. The basic station network of the China Earthquake Administration's geomagnetic station has been using fluxgate magnetometers to detect changes in the geomagnetic field in seismic danger zones since 2003~\cite{Wang-2008-J,Wang-2012-J}. This method can be used to detect earthquake anomaly precursors in time and provide basic information for accurate prediction of earthquake occurrence. Meanwhile, the observed geomagnetic field strength, inclination, and declination can be adopted to study the Earth's basic magnetic field and deep underground structures~\cite{Prouty-2013-J}. 

In addition, the high-quality observation data of the ionosphere, the substorm current, and the very-low-frequency (VLF) magnetic field of the measured region can be used to study the space weather~\cite{Liu-2017-J4}. For the particularity of observations, the fluxgate magnetometer array system can realize the Network functions such as communication access, automatic identification, monitoring \& management, and data collection for field mobile observation equipment. According to the temporal and spatial distribution characteristics of the geomagnetic field, each three-component fluxgate magnetometer which is arranged in the basic network channels, is connected to the monitoring center through the wireless or wired Internet network, and a large amount of geomagnetic raw observation data is acquired after a period of recording to achieve reliable and real-time monitoring of the geomagnetic data. This system plays an important role in the observation platform for short-period geomagnetic. However, the observation interruption will occur due to some factors such as the observation site, power supply, communication conditions, etc., and its stability needs to be further strengthened. Hence, a Superconducting Quantum Interference Device (SQUID) could be considered to measure the variation of the three-component magnetic field to obtain a higher sensitivity~\cite{Korepanov-2006-J,Chave-1995-J}.
\begin{figure}[htb]
\centering
\includegraphics[width=1.0\linewidth]{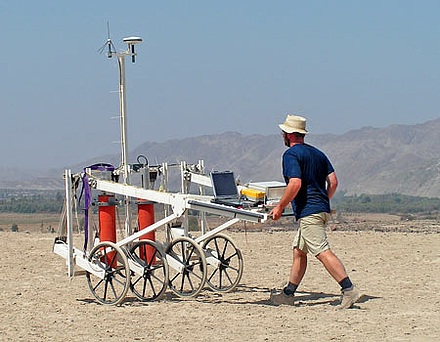}
\caption{JESSY SMART ground tensor magnetic gradient measurement system developed by IPHT~\cite{Linzen-2007-J}.}
\label{LTS SQUID}
\end{figure}

For the mobile measurement, the Institute for Photonic Technologies (IPHT) of Jena, Germany, adopted a low temperature superconducting (LTS) SQUID to develop a ground tensor magnetic gradient measurement system, named JESSY SMART~\cite{Linzen-2007-J}, as shown in Fig.~\ref{LTS SQUID}. The JESSY SMART is a novel ground tensor magnetic gradient measurement system for fast three-dimensional (3D) geomagnetic mapping of hidden subsurface anomalies. It can localize smallest Earth's magnetic field gradient inhomogeneities in a depths of up to approximate 10 meters. The sensors are mounted onto a cart which can be moved attached to a motor car or alternatively pushed along manually. It has the following advantages when compared with conventional geomagnetic methods: 1) highest possible magnetic field resolution (sensor 7 fT/cm); 2) fast mapping of measuring area (up to 7 acre and accordingly 3 ha/hour); 3) very accurate position and spatial resolution of data in cm range); 4) measurement in difficult ground, environment, and climate conditions possible; 5) nondestructive measuring and a high planning reliability.    
\begin{figure}[htb]
\centering
\includegraphics[width=1.0\linewidth]{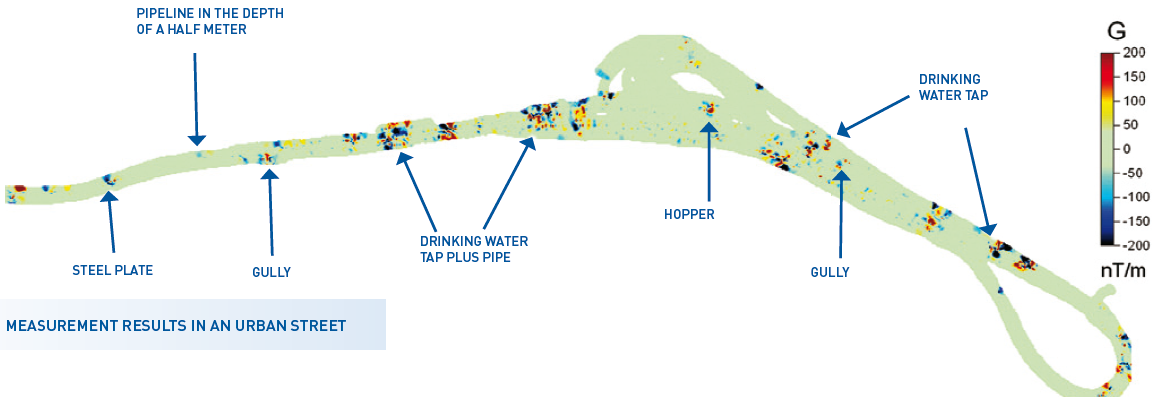}
\caption{Investigation and clearing of UXO before road construction using the JESSY SMART~\cite{Meyer-2009-J}.}
\label{Result}
\end{figure}

Figure~\ref{Result} shows the measurement results from a 400-meter length area in an urban street of the city center, Plauen, Germany. The mission is to investigate and clear the UXO before road construction. To minimize the road traffic and construction work disturbing influence, the mission was implemented at night time. The magnetogram shows a couple of extended anomalies which correspond to various metallic objects assembled in refilled pits, a typical residual of the very first simple clearance procedures carried out directly before the construction. The magnetic anomalies and the corresponding localization of the magnetic objects can be visualized clearly, which demonstrates the effectiveness of the system.

In 2017, the Strasbourg Institute of Globe Physics~\cite{Gavazzi-2017-J} developed a fluxgate based mobile measurement system for georeferenced magnetic measurements at different scales, i.e., from sub-metric measurements on the ground to aircraft-conducted acquisition through the wide range offered by unmanned aerial vehicles (UAVs), with a precision in the order of 1 nT. Such equipment can be used for different kinds of applications, such as structural geology, pipes, and UXO detection, archaeology. Likewise, Jilin University proposed a novel noise reduction method and developed a corresponding magnetic resonance three-component noise reduction device~\cite{Lin-2014-P,LinT-2017-J}. However, this method is in the laboratory stage, and its practicability for field application still needs to be further validated.

\subsection{Wells vector magnetic measurement}
Wells vector magnetic survey technique is used for census exploration of magnetite deposits or poly-metallic deposits containing ferromagnetic minerals. It is a geophysical prospecting method for the development and extension of the ground magnetic surveys in space, based on the study of the magnetism of rocks and ore bodies. Besides, this technique has a unique advantage in the exploration of deep magnetic deposits, and it can solve some problems that the surface magnetic survey cannot solve, for instance, the depth position of the ore tail and the top of the mine. The wells magnetic survey can avoid the human-made interferences and the heterogeneity of shallow magnetic sources, and the vertical resolution is relative high. Because of these advantages, the three-component magnetic survey in the well has become an effective method for surveying every hole of the magnetite deposit in the general survey~\cite{Chen-2008-J}.

The three-component magnetic survey in the well is carried out along the drilling direction of the borehole. Through measuring the $x$, $y$, and $z$ components of the magnetic field at different depths in the well, the magnetic anomaly components $\Delta x$, $\Delta y$, and $\Delta z$ can be calculated to judge and evaluate the blind magnetic ore beside the well, the blind ore at the bottom of the well, the shape of the ore body, the size of the mineral, etc~\cite{Silva-1981-J,Parker-1979-J,Zhang-2007-J}. The three components $\Delta x$, $\Delta y$, and $\Delta z$ can also be synthesized into an total intensity anomaly $\Delta T$. 

In practical applications, aiming to facilitate the qualitative study of magnetic anomalies, $\Delta T$ is divided into a vertical component $\Delta Z$ and a radial component $\Delta H$, and the $\Delta H$ can be further divided into a horizontal component which is perpendicular to the direction of the ore body and a horizontal component which is parallel to the direction of the ore body. In this case, we can obtain the relations between the magnetic anomaly strength and the direction of the ore body. Through the analysis of the parameters such as the magnetic vertical component and the radial component of the magnetic anomaly area, we can infer whether there are ore bodies at the bottom of the well or besides the well.
\begin{figure}[htb]
\centering
\includegraphics[width=1.0\linewidth]{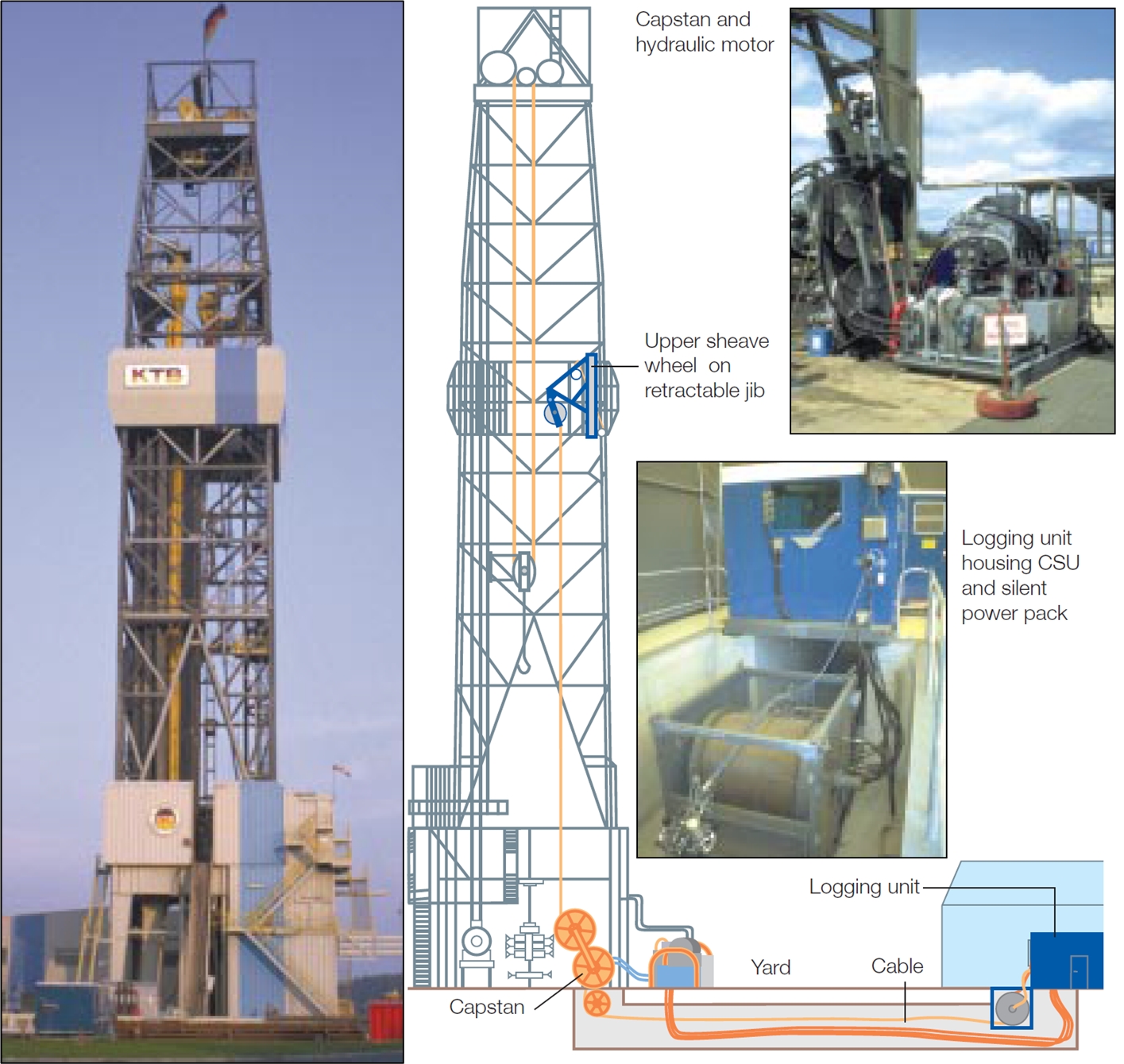}
\caption{Overall view of the KTB drilling system and the corresponding module structure~\cite{Bram-1988-J}.}
\label{KTB}
\end{figure}

In the mid-1960s, Wang et al.~\cite{Wang-2009-J} developed China's first three-component magnetic survey system for transistor wells. This system can directly measure the three-component magnetic field value of the vertical coordinate system in the well, using a gravity orientation with two degrees of freedom. When compared with the degree of freedom (DOF) gravity orientation system developed in Sweden at that time, it has obvious advantages such as high measurement accuracy, real-time measurement, etc.

In the 1990s, the German Continental Drilling Program (KTB) has made great progress in data processing and inversion interpretation of magnetic surveys in boreholes~\cite{Emmermann-1997-J}. Fig.~\ref{KTB} shows the overall KTB drilling system~\cite{Bram-1988-J}, in which a combination of a closely spaced surface gravity survey with a high-resolution helicopter aeromagnetic survey as well as borehole gravity and magnetometer measurements allowed a detailed 3D modeling of the anomalies at the KTB drill site. To be specific, in the KTB main hole, numerous fluxgate magnetometers are equipped in the vertical drilling system (VDS), collecting the magnetic field data. The magnetic measurements are always carried out at depths from 300 m to 6000 m, and the errors are less than 10 nT in total after data processing including special correction, sophisticated calibration, etc. For a better resolution of the magnetic anomaly, a detailed helicopter survey is carried out with the drill site in its center. Radiometric and electromagnetic data are also collected. In this case, an almost complete coverage and high resolution of the magnetic field in the measuring level is obtained, yielding an excellent data set for analyzing the geological structure and supporting surface geological mapping.

\begin{figure}[htb]
\centering
\includegraphics[width=1.0\linewidth]{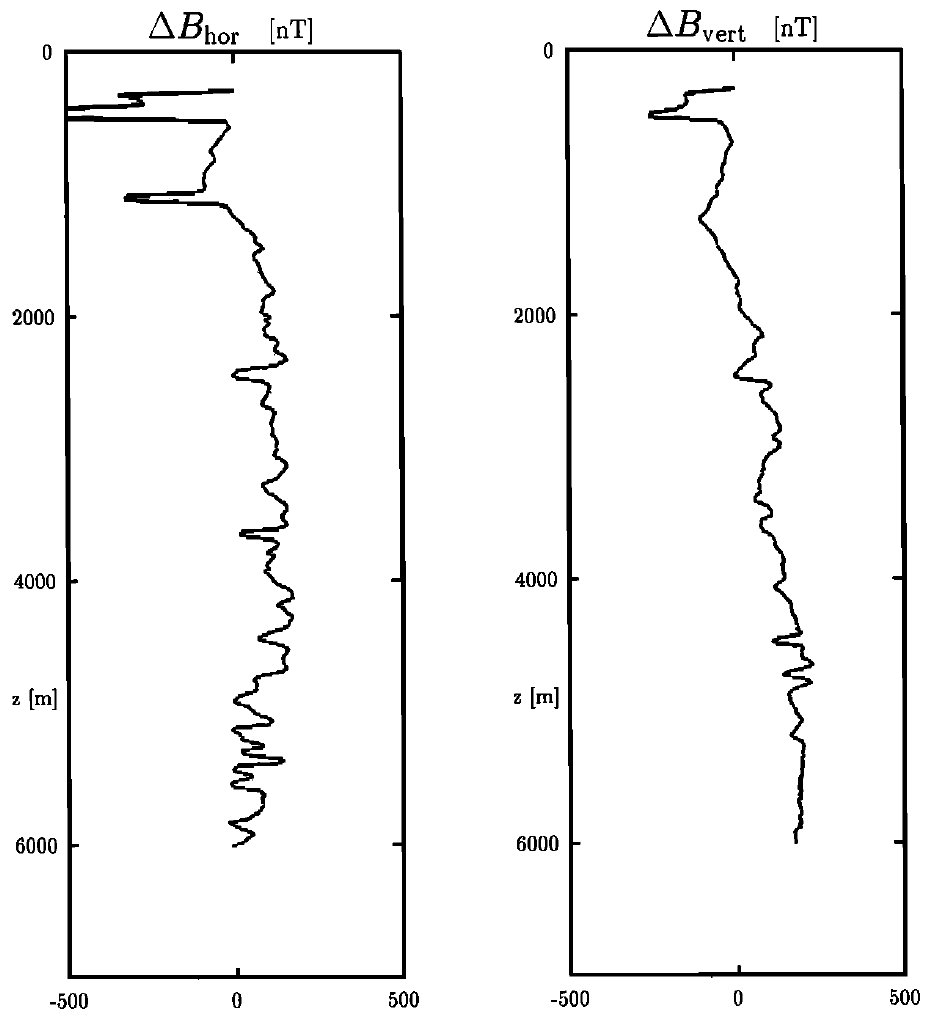}
\caption{The trend of the horizontal and vertical component of the magnetic field in the KTB main drill hole~\cite{Bosum-1997-J}.}
\label{Result3}
\end{figure}

A surprising result using the KTB is an anomalous vertical gradient of the geomagnetic field which was detected in the lower part of the borehole down to at least 6000 m. To emphasize the general trend of the magnetic field in the borehole, the data were processed with a running 40 m low-pass median filter. The results are shown in Fig.~\ref{Result3}. Without any magnetized bodies at depth, a vertical gradient of the total magnetic field of about 22 nT/km should be expected from the local international geomagnetic reference field (IGRF). The source bodies in the upper part of the borehole produce even a decreasing field in the lower part of the drill hole. However, gradients of up to 200 nT/km have been measured. The average gradient below 3000 m is 60 nT/km. Below 6000 m the trends and gradients of the magnetic field could not be determined because of the low accuracy of the orientation tool.

Moreover, the United States, Sweden, and Canada have made important contributions to the design of magnetic measuring instruments, data processing, etc. For instance, Kuhnke et al.~\cite{Kuhnke-1992-C} designed a borehole magnetometer, analyzed the possible problems of the magnetometer during operation in the well, and developed different types of instrument systems based on the temperature resistance characteristics. Bosum et al.~\cite{Bosum-1997-J} analyzed the characteristics of borehole magnetic survey data, and Worm~\cite{Worm-1993-J} conducted rock magnetic simulation studies by combining with KTB borehole magnetic survey data. Leonardi et al.~\cite{Leonardi-1998-J} adopted KTB magnetic survey data to study the anisotropy of the crustal structure, and obtained some fractal features.

In the early 21st century, a new three-component magnetometer, dubbed Goettinger borehole magnetometer which has been successfully used to measure in the Outokumpu Deep Drill Hole (OKU R2500, Finland), the repeatability of the magnetic field vector is 0.8$^{\circ}$ in the azimuthal direction, 0.08$^{\circ}$ in inclination and 71 nT in magnitude~\cite{Virgil-2015-J}. Moreover, a reoriented vector borehole measurement was carried out using the Goettinger magnetometer at two sites during Integrated Ocean Drilling Program (IODP) Expedition 330 to the Louisville Seamount Chain~\cite{Ehmann-2015-J}. Likewise, Chongqing Geological Instrument Factory developed a high-precision three-component magnetometer, named GJCX-1~\cite{Liu-2013-J}. The orientation differences of $x$ and $y$ direction are both less than 100 nT, the orientation difference of $z$ direction is less than 50 nT. Shanghai Institute of Geosciences developed a high-precision three-component logging tool, named JCC3-2A, in which the three-axis giant magnetoresistive sensor is used as the magnetic sensitive element, and the three-axis gravity acceleration sensor is used as the directional element. The sensitivity of the magnetic component is 40 nT, and the accuracy is better than 100 nT~\cite{Yuan-2016-J}.

Consequently, the wells three-component magnetic survey technique has the advantage of a large radial detection range, which provides an effective tool for dividing the magnetic rock or ore body on the borehole profile, the magnetic minerals besides or below the exploration well, and the blind ore bodies associated with the magnetic minerals. To a certain extent, it is possible to perform depth calibration on the inversion of the ground magnetic survey, solving the detection work that cannot be completed by the ground magnetic survey technique, which implies that the wells vector magnetic survey an effective method for deep ore body detection. However, due to the limitations of borehole diameter, well temperature, and the size \& volume of the instrument itself, the accuracy of the magnetic sensor used in the wells magnetic survey is lower than that of the ground magnetic survey, and the overall observation accuracy is also lower than the surface magnetic survey by an order of magnitude. Hence, it is of great significance to develop a higher-precision three-component magnetic measuring instrument in the well.

\subsection{Marine vector magnetic measurement}
Marine magnetic survey is the main approach to obtain the information on the distribution and variation characteristics of the geomagnetic field in the ocean area, and it is also one of the important cases of marine engineering surveys and military hydrographic surveys~\cite{Boyce-2004-J}. In the 1950s, Vacquier et al.~\cite{Vacquier-1951-J} adopted a fluxgate magnetometer to measure the geomagnetic field in the three oceans. Since the 1980s, the international geoscience community has begun to implement seafloor geomagnetic observations, deploying dozens of seafloor observation stations to conduct oceanic and geomagnetic surveys, and carrying out research on underground deposits and geodynamics. Afterwards, the earth-based electromagnetic measurement method based on artificial excitation appeared, e.g., electromagnetic method of ocean controlled source~\cite{Bochu-2011-J}. 

Up to now, the marine magnetic survey mainly employs an optical pump magnetometer or an Overhauser magnetometer for the underwater magnetic survey~\cite{Weiss-2007-J,Zong-2015-J,Dong-2018-J}. To suppress the influence of waves, the magnetic probes are usually measured at a certain depth below the surface of the ocean, using a drag-and-drop measurement approach. In 1997, Seama et al.~\cite{Seama-1997-J} developed a vector magnetometer for deep tow detection. To determine the position of the tow body in the deep ocean, an inertial navigation system and GPS, short baseline acoustic measurement and pressure measurement were adopted. Gee and Cande~\cite{Gee-2002-J} developed a vector magnetometer system with a speed of 10 $\sim$ 12 knots. The test results show that the device can determine the horizontal and vertical components with an accuracy of 30 nT $\sim$ 50 nT, which is of great significance for magnetic survey applications in low latitude areas.

To further identify the magnetic anomalies caused by short-polarity events or other local magnetic bodies, Blakely et al.~\cite{Blakely-1973-J} developed a magnetic vector measurement system. Besides, Horner and Gordon~\cite{Horner-2003-J} adopted spectral analysis methods to identify anomalies in the seabed magnetic bands, and they found that the aeromagnetic survey data had better results than shipboard total field survey data. 

\begin{figure}[htb]
\centering
\includegraphics[width=1.0\linewidth]{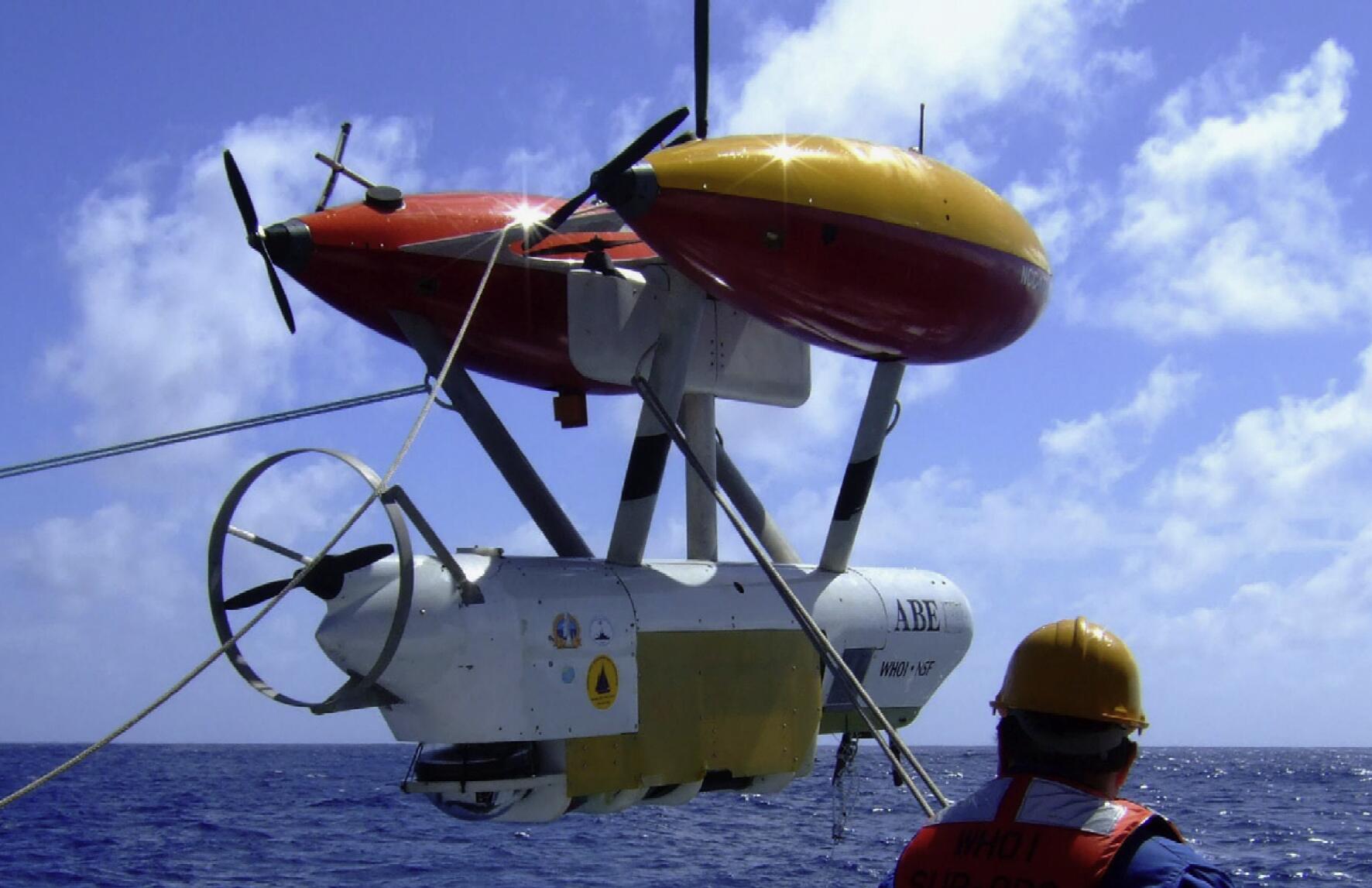}
\caption{Autonomous underwater vehicle ABE~\cite{German-2008-J}}
\label{Seaquest}
\end{figure}

In 1995, the Woods Hole Oceanographic Institute developed the world's first unmanned submersible autonomous benthic explorer (ABE) for marine magnetic survey~\cite{German-2008-J}, which is mainly composed of conductivity probe, temperature probe, depth gauge, camera and magnetometer, as shown in Fig.~\ref{Seaquest}. The ABE is a robotic underwater vehicle used for exploring the ocean to depths of 4,500 meters. It was the first AUV used by the U.S. scientific community. ABE is often used in tandem with Alvin or Jason surveying large swaths of ocean floor to determine the best spots for close-up exploration. ABE is designed to perform a pre-programmed set of maneuvers, using its five thrusters to move in any direction, hover, and reverse. The AUV excels at surveys of the shape of the seafloor, its chemical emissions, and its magnetic properties. ABE is particularly valuable in rugged terrain. On-board sensors tell the vehicle how deep it is and how far it is off the ocean floor, and the AUV calculates its horizontal position by contacting a system of acoustic beacons (transponders) set out in fixed locations~\cite{Yoerger-1991-C}. In 2014, the 715-research institute developed the RS-HC3 ocean tensor magnetic gradiometer which dynamic range is -100000 nT $\sim$ 100000 nT, and this system also has got a preferable result.

\begin{figure}[htb]
\centering
\includegraphics[width=1.0\linewidth]{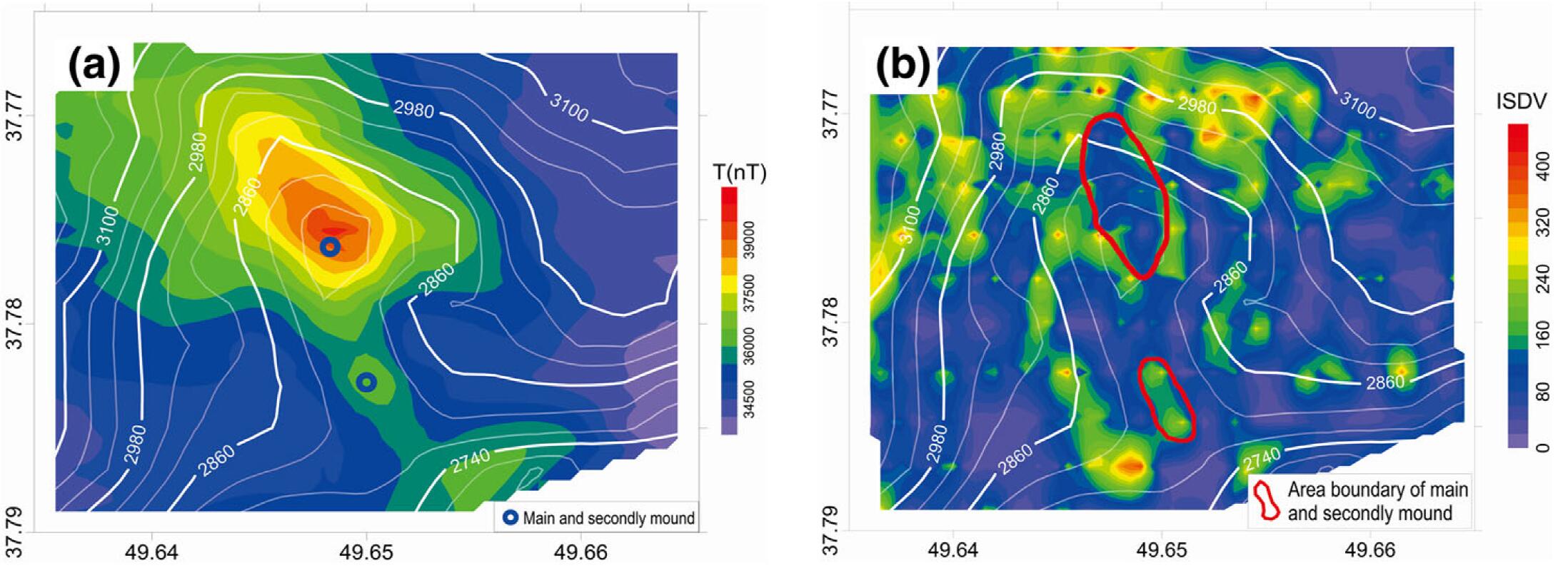}
\caption{Fault inference in the Longqi hydrothermal area based on near-bottom magnetic data collected by the ABE~\cite{Tao-2017-J}. (a) Near-bottom magnetic field; (b) The spatial differential vector (ISDV) distribution.}
\label{Result4}
\end{figure}

Figure~\ref{Result4} shows the fault inference in the Longqi hydrothermal area based on near-bottom magnetic data collected by the ABE 200 dive. The blue circles point location of the mounds and the solid red lines indicate the trap areas. There is an obvious high magnetic anomaly on the top of the main mound and only a weak positive anomaly on the secondary mound, because the anomaly resolution decreases with increasing depth. In addition, the corresponding ISDV values are well-defined for the main and secondary mounds with the values showing a well-defined trap area and a smaller secondary trap.

\subsection{Airborne vector magnetic measurement}
The resource conditions and environmental conditions in many areas of the earth are complex, such as swamps, forests, deserts, and mountain areas, where are inconvenient for personnel equipment to access~\cite{Dransfield-2003-J,Huang-2008-J}. The airborne electromagnetic detection system can overcome the impact of the complex ground environment. It has the characteristics of low cost, fast survey speed, wide survey area, and strong versatility, and plays an important role in geophysical exploration~\cite{Liu-2019-J6,Liu-2019-J8}. Before the 21st century, the airborne magnetic survey technique is mainly based on the measurement of total field strength or gradient. As the continuous development of magnetic exploration theories and methods, the airborne survey technique has been promoted from the original total field strength or gradient measurement to vector (three-component or gradient tensor) measurement~\cite{Hardwick-1984-J,Salem-2001-J}. 

In 2003, the Australia BHP Billiton Petroleum Company developed a fluxgate aviation three-component magnetic measurement system, and conducted flight tests in the Rocky Strip iron-bearing construction area of Western Australia~\cite{Dransfield-2015-J}. After attitude correction, three-component geomagnetic field data and attitude were obtained, in which the corrected noise level is 50 nT $\sim$ 100 nT. In 2011, Munschy et al.~\cite{Munschy-2011-J} installed two fluxgate magnetometers on the tail of the Maule MX7 small aircraft, and conducted flight tests in the Vosges area. After magnetic compensation, the magnetic field data of two horizontal components were obtained, and the vertical component was calculated by combining the measurement results of the total field. After mapping the magnetic field distribution in this area, it is basically consistent with the actual geological conditions. In the same year, the China Aero Geophysical Survey and Remote Sensing Center (AGRS) developed an airborne three-component magnetic survey system, which is composed of a fluxgate magnetometer and an INS/GPS strap-down inertial navigation module~\cite{Ge-2019-J1}. In addition, a verification test was conducted by mounting this system to a fixed-wing aircraft~\cite{Lin-2017-J}. 

At the beginning of the 21st century, another airborne vector magnetic survey technique: aeronautical full-tensor magnetic gradient measurement has become a hot spot for geophysicists in various countries. This technique refers to measure the spatial change rate of the three field components of the geomagnetic field vector along three mutually orthogonal axes, and there are nine elements in total~\cite{Liu-2020-J1}. Its outstanding advantages include the invariant contour plot calculated from the gradient tensor is easy to explain, the dipole tracking algorithm can be used to accurately determine the depth and horizontal position of the magnetic dipole, etc. Hence, the high-precision 3D positioning of underground magnetic geological bodies and ore bodies can be realized through full-tensor magnetic gradient measurement data, and their spatial distribution information can be obtained~\cite{Liu-2020-J2}. 

In 2003, the U.S. Geological Survey (USGS) conducted proof-of-principle tests of the tensor magnetic gradiometer system (TMGS) at the Strategic Environmental Research and Development Program's (SERDP) UXO Standardized Test Site at the Yuma Proving Ground (YPG), Arizona~\cite{Smith-2004-C}. The noise of this system in 0.001 Hz $\sim$ 100 Hz band is typically less than 0.1 nT, and the thermal stability is about 0.2 nT deg$^{-1}$ for a field of 60000 nT~\cite{Narod-1990-J}. The objective of these tests was to assess the potential of this system for UXO applications and to help us formulate design parameters for an improved TMGS designed specifically for UXO. Although the current system was not designed with UXO applications in mind, it detected the majority of the buried targets in the Calibration Area at YPG. Its performance rivaled that of the cesium vapor companion system, and it showed true tensor gradiometric performance over a selected target. 
\begin{figure}[htb]
\centering
\includegraphics[width=1.0\linewidth]{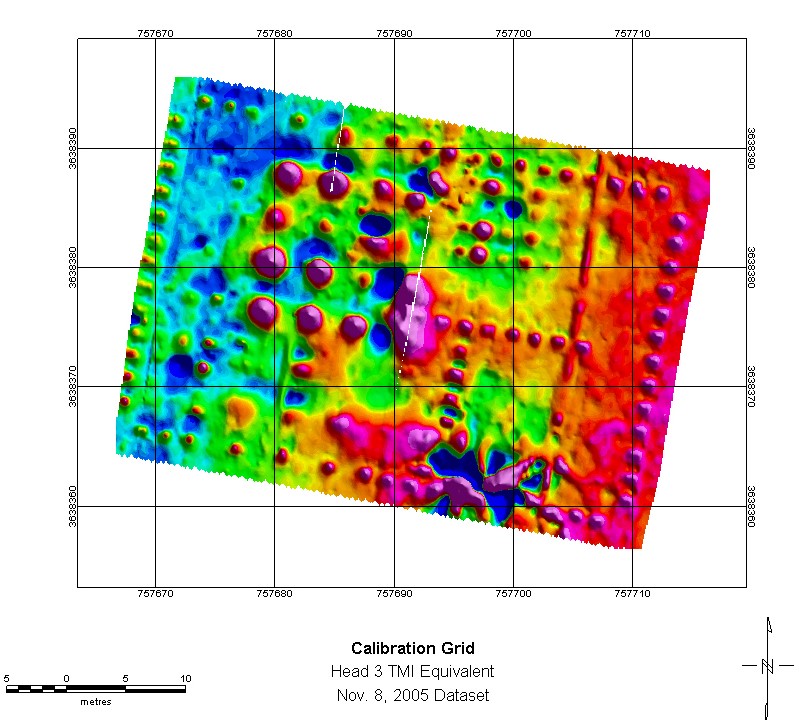}
\caption{The tensor invariant map over a 60 mm mortar round buried 0.25 m deep of the YPG measuring by the TMGS~\cite{Smith-2004-C}.}
\label{Result1}
\end{figure}

Figure~\ref{Result1} shows a tensor invariant map over a 60 mm mortar round buried 0.25 m deep of the YPG measuring by the TMGS. The array was towed over the grid at an average height of 0.3 m, and most of the measured magnetic gradients greatly exceed 2 nT/m. The tensor invariant maps have been despiked and corrected for GPS lag, and system timing latency errors. The tensor anomaly is positive and highly peaked over the target at the center of the grid. The positive anomalies at the edges of the map correspond to adjacent off-grid targets.

Jilin University constructed a magnetic full-tensor gradiometer, which utilizes four fluxgates arranged on a planar cross structure, and a single, triaxial, spherical feedback coil assembly~\cite{Sui-2014-J}. In this arrangement, one of the fluxgates is used as a reference, controlling the currents through the feedback coils. Since the fluxgates are working in the near-zero magnetic field environment, the magnetic tensor gradiometer is stable and of an improved accuracy. However, due to the low sensitivity of the fluxgate magnetometer, the baseline distance between the sensors is very large when performing gradient measurement, and thus, the requirements on the system structure are relative high, and the measurement accuracy is difficult to guarantee. Consider for this case, the SQUID based vector magnetic sensor with higher sensitivity has become the first choice for the magnetic gradient tensor measurement system. 
\begin{figure*}[htb]
\centering
\includegraphics[width=0.87\linewidth]{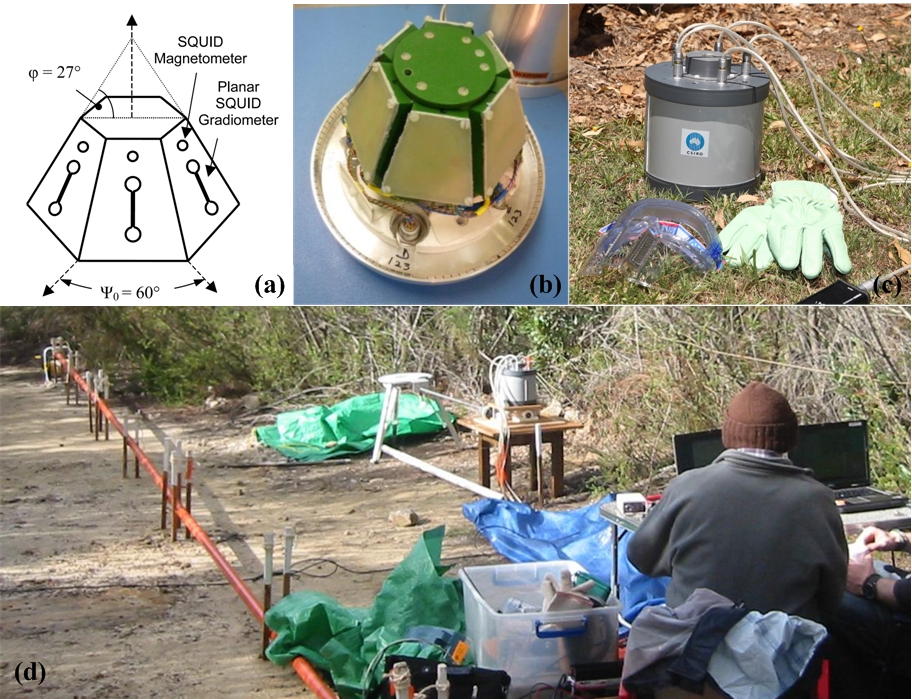}
\caption{HTS full-tensor magnetic gradiometer developed by CSIRO~\cite{Keenan-2010-J}. (a) The magnetic probe's structure; (b) The magnetic probe; (c) The overall system; (d) System experiment on ground.}
\label{UXOMAG}
\end{figure*}

In 2005, the U.S. Oak Ridge National Laboratory (ORNL) developed an aeronautical full-tensor magnetic gradient measurement system which is composed of high temperature superconducting (HTS) SQUIDs and an aeronautical geophysical platform~\cite{Gamey-2004-C}, and implemented a ground experiment for this system. The system adopts eight HTS-SQUID sensors which are all fixed on a bracket, forming a full-tensor probe through a reasonable layout.  
\begin{figure*}[htb]
\centering
\includegraphics[width=0.87\linewidth]{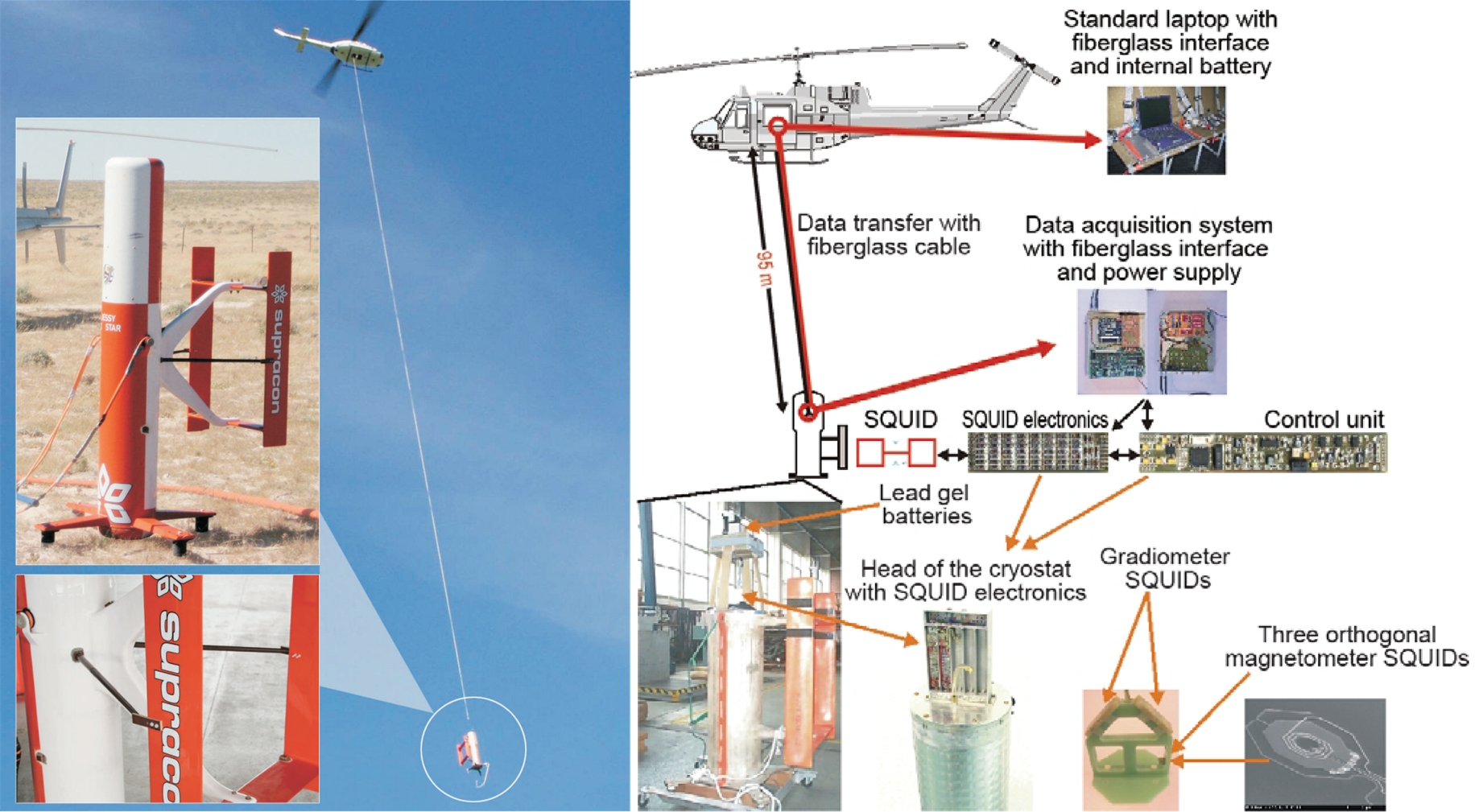}
\caption{Setup of the airborne SQUID system developed by IPHT~\cite{Stolz-2006-J}.}
\label{1-generation}
\end{figure*}

Likewise, the Commonwealth Scientific and Industrial Research Organization (CSIRO) developed an HTS-SQUID based areo full-tensor magnetic gradient measurement system, namely, UXOMAG, as shown in Fig.~\ref{UXOMAG}. The UXOMAG is a mobile magnetic tensor gradiometer prototype system capable of not only detecting, but also locating and classifying UXO. This system can measure all five independent magnetic tensor components allowing the user to discriminate between harmless metallic debris or shrapnel and a potentially lethal UXO~\cite{Keenan-2010-J}. It uses multiple CSIRO's HTS planar SQUID gradiometers configured unique geometry to determine the magnitude of the full-tensor. Referencing magnetometers are used to improve common mode rejection. The UXOMAG's high sensitivity has the potential to detect 40 mm calibre UXO at a distance of up to 4 m away. 

There are also some other SQUID based full-tensor magnetic gradient measurement systems, e.g., the prototype developed by Jilin University~\cite{Shen-2016-J}. For this system, six first-order planar-type SQUID gradiometers mounted on the six sloping facets of a hexagonal pyramid, which can form the probe and determine all independent components of the magnetic gradient tensor with the help of triaxial SQUID magnetometers placed near the probe. The gradiometer sensitivity is about 100 pT/m. The Shanghai Institute of Microsystem and Information Technology (SIMIT) and the Institute of Remote Sensing and Digital Earth (RADI) from the Chinese Academy of Sciences jointly developed a prototype of HTS-SQUID based areo full-tensor magnetic gradient measurement system, and conducted a flight measurement test in Inner Mongolia of China. The gradiometer sensitivity is up to 50 pT/m. However, these two systems are still in the experimental stage.

Up to now, the most commonly used and accepted practical aero full tensor magnetic gradient system is developed by IPHT. The IPHT won the "Outstanding Achievement Award" in the category Mining Research for their ground breaking research in the field of world's first full-tensor airborne magnetic gradiometer, known by its synonym "JESSY STAR". Fig.~\ref{1-generation} illustrates the overall system. The JESSY STAR system has surpassed all conventional surveying technology in performance and, in addition, provided the exploration community with magnetic data never measured before. It also utilizes the ultimate sensitivity of LTS-SQUID sensors~\cite{Zakosarenko-2003-J}. The JESSY STAR system include six gradiometers and three magnetometer channels, to compensate for sensor motion noise. Further, high spatial resolution is provided by a differential GPS system using SBAS and an inertial unit (INU), which are synchronized with the magnetic gradient data, The INU data is used for motion compensation. The data acquisition system, incorporate a small-sized 15 channels 24 bit analogue digital converter unit. To date, JESSY STAR has flown hundreds of hours in both helicopter mode and fixed wing plane configuration, and it has transformed exploration from "flying blind" to "seeing clearly". This new generation system is capable of detecting minerals and precious metals that were deemed "undetectable" before.

The magnetic measurements with a UAV are ideal for filling the gap between the ground and airborne magnetic surveying. For several years, there has been a very important development of magnetic measurement with UAVs by using the scalar or vector magnetometers. However, for now, although the vector magnetic measurement has numerous advantages, the technological limitations in sensor manufacturing and unwanted magnetic fields will corrupt the measurements of three-axis magnetometers, especially in the airborne or satellite vector magnetic survey. Since the three-axis magnetometer (take the fluxgate as an example) is not an absolute instrument, it has to be calibrated and compensated. 

Generally, there are two different principles for calibrating a magnetometer. In a vector calibration, the output of the vector magnetometer is compared with the known magnetic field vector which is applied to the instrument. The problem of an in-flight vector calibration is of course that the applied magnetic field vector is in principle unknown. In a scalar calibration, only the intensity of the magnetic field is used, but not its direction. However, the scalar intensity is known from the second magnetometer, e.g., the Overhauser scalar magnetometer. Calibration of the vector magnetometer is based on the following points: 1) use pre-flight values for the temperature dependence; 2) use scalar in-flight calibration for the estimation of the nine intrinsic parameters of the vector magnetometer; 3) use vector in-flight calibration for the simultaneous estimation of the remaining three calibration parameters together with a spherical harmonic model of the Earth's magnetic field. More details can be refereed to literature~\cite{Olsen-2001-J,Munschy-2007-J}.

Likewise, Gavazzi et al.~\cite{Gavazzi-2016-J} implemented a case study on the aerial military base BA112 shows the usefulness of the instruments for the detection of underground pipes, UXO, and archaeological remains. Maire et al.~\cite{Le-2020-J} proposed a new method for rapid mapping and upscaling from the field to a regional scale using a UAV and a fluxgate magnetometer. However, to obtain accurate aeromagnetic data, the compensation of the magnetic effects of the UAV is a challenge. Most research on aeromagnetic compensation has mainly focused on the suppression of the magnetic interference generated by aircraft maneuvers and the on-board electronic equipment. Amount of compensation methods, such as ridge regression~\cite{Guerard-1984-J}, truncated singular value decomposition~\cite{Mingqiu-1997-J}, recursive least-squares~\cite{Ding-2006-J}, partial least-squares~\cite{Sohn-2008-J}, support vector methods~\cite{Wu-2017-J}, c-k class estimation~\cite{Noriega-2017-J}, and principal component analysis~\cite{Wu-2018-J} have been proposed one after another, ensuring the high-quality of measured aeromagnetic data. 

The inertial measurement units (IMU) is an affordable instrument for the determination of orientation. The fluxgate sensors embedded in the system are affected by nonidealities that can be greatly compensated by proper calibration, by determining sensor parameters, such as bias, misalignment, and sensitivity/gain. For instance, an online calibration method for a three-axial magnetometer using a 3D Helmholtz coil is proposed in the literature~\cite{Beravs-2014-J}. The magnetometer is exposed to different directions of the magnetic field created by the 3D coil. The parameters are estimated by using an unscented Kalman filter. The directions are calculated online by using a sensor parameter covariance matrix. 

\begin{figure}[htb]
\centering
\includegraphics[width=1.0\linewidth]{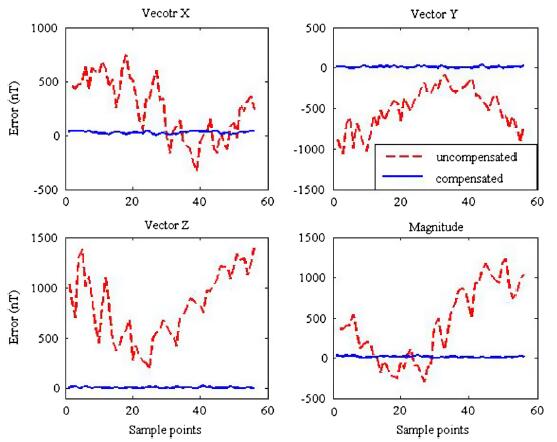}
\caption{Uncompensated and compensated results~\cite{Zhang-2016-J}.}
\label{Compensation}
\end{figure}

In addition, ferromagnetic interferential field from platforms is one of the most dominating error sources for magnetometer. For magnetic vector and gradient tensor measurement, what is cared about most is the effect of compensating magnetic field vector. For instance, a novel compensation method has been proposed in literature~\cite{Zhang-2016-J}, in which attitude information is used to set up a vectorial compensation model and least square method is used to estimate parameters. Fig.~\ref{Compensation} shows the uncompensated and compensated results. It can be seen that the measurement errors of geomagnetic field vectors and magnitude decrease remarkably after compensation.

\subsection{Satellites vector magnetic measurement}
In 1958, the Soviet Union launched the first satellite SPUTNIK-3 to measure the geomagnetic field, which foreshadowed the beginning of the satellites magnetic survey technique~\cite{King-1958-J,May-1962-J}. The fluxgate magnetometer was installed on this satellite, however, since the direction of the magnetometer cannot be accurately determined, only total field strength data were obtained. Since then, a series of magnetic survey satellites carried total field magnetometers were launched by the Soviet Union, including proton precession magnetometers or optical pump magnetometers. Likewise, they cannot be called vector magnetic survey. With the development of space magnetic field measurement technology, there have been numerous professional satellite programs dedicated to the mapping of the Earth's intrinsic geomagnetic, such as the U.S. MAGSAT satellite program~\cite{Langel-1980-J,Hulot-2002-J}, Denmark's ${\O}$rsted satellite program~\cite{Duret-1996-J,Rj-2000-J}, Germany's CHAMP satellite program~\cite{Reigber-2002-J,Huang-2017-J}, Argentina's SAC-C satellite program~\cite{zhou-2009-J}, and European Space Agency's (ESA's) Swarm~\cite{Olsen-2013-J}.

The MAGSAT spacecraft was launched into a twilight, sun-synchronous, orbit with inclination 96.76$^{o}$, perigee 352 km and apogee 561 km~\cite{Langel-1982-J}. The Cesium vapor scalar and fluxgate vector magnetometers together measured the field magnitude to better than 2 nT and each component to better than 6 nT~\cite{Langel-1982-J1}. Two star cameras, a high-accuracy sun sensor and a pitch axis gyro provided the 10-20 arc-second attitude measurements necessary to achieve this accuracy. The magnetometers were located at the end of a boom to eliminate the effect of spacecraft fields. An optical system measured the attitude of the vector magnetometer and sun sensor (at the end of the boom) relative to the star cameras (on the main spacecraft)~\cite{Acuna-1978-J}. The data are available in several formats from the National Space Science Data Center and are undergoing analysis by a team of investigators~\cite{Langel-1985-J}. In addition, An et al.~\cite{An-1998-J,An-1998-J1} employed the method of spherical cap harmonic analysis to derive a spherical cap model based on the MAGSAT data set, and described the three-dimensional structure over Asia and Europe region, respectively.   
\begin{figure}[htb]
\centering
\includegraphics[width=0.9\linewidth]{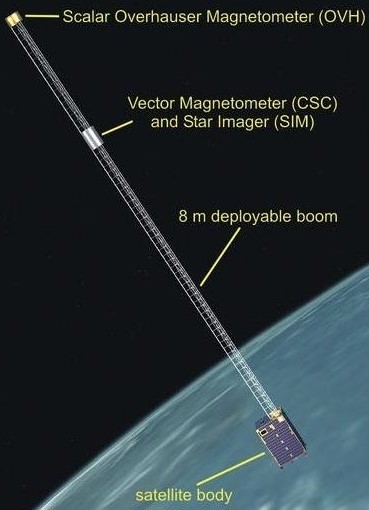}
\caption{Illustration of the ${\O}$rsted spacecraft in orbit~\cite{Olsen-2003-J}.}
\label{Orsted}
\end{figure}

The ${\O}$rsted satellite, named after the Danish scientist Hans Christian ${\O}$rsted, is the first satellite mission since MAGSAT designed for high-precision mapping of the Earth's magnetic field~\cite{Olsen-2001-J}. It was launched with a Delta-II rocket from Vandenberg Air Force Base into a near-polar orbit. As the first satellite of the "International Decade of Geopotential Research", the satellite and its instrumentation have been a model for other present and forthcoming missions like CHAMP and Swarm~\cite{Olsen-2003-J}. The ${\O}$rsted spacecraft is composed of two main instruments, including an Overhauser magnetometer and a compact spherical coil based triaxial fluxgate magnetometer, as illustrated in Fig.~\ref{Orsted}. The objective of the Overhauser magnetometer is to measure magnetic field scalar values with an absolute measurement error better than 0.5 nT, an dynamic range from 16000 nT to 64000 nT, and a sampling rate of 1 Hz~\cite{Neubert-2001-J}. The fluxgate magnetometer which measures magnetic field vectors at an absolute measurement error less than 1 nT, an dynamic range from -65536 nT to 65536 nT, and a resolution better than 0.25 nT. Further, the collected vector data is calibrated using the absolute intensity measured by the Overhauser magnetometer. After calibration, the agreement between the two magnetometers is better than 0.33 nT~\cite{Nielsen-1995-J}.
\begin{figure}[htb]
\centering
\includegraphics[width=0.9\linewidth]{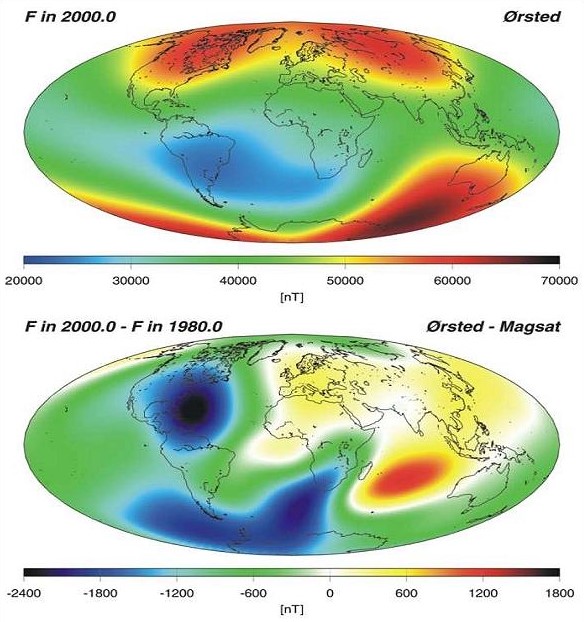}
\caption{Illustration of the geomagnetic field of two periods (1980 and 2000)~\cite{Olsen-2000-J}.}
\label{Result2}
\end{figure}

Figure~\ref{Result2} illustrates the geomagnetic field of two periods (1980 and 2000) recorded by the ${\O}$rsted satellite. The top figure displays a color-coded global image of the magnetic field strength in the year 2000 as modelled from the ${\O}$rsted data. The scale spans from 20000 nT $\sim$ 70000 nT. The bottom field displays the differences between ${\O}$rsted results from 2000 and results from the MAGSAT mission. The scale spans here from - 2400 nT to + 1800 nT. The changes in field strength over the 20 years span between the two missions are mostly negative and ranges up to almost 10\% of the total field.
\begin{figure}[htb]
\centering
\includegraphics[width=0.9\linewidth]{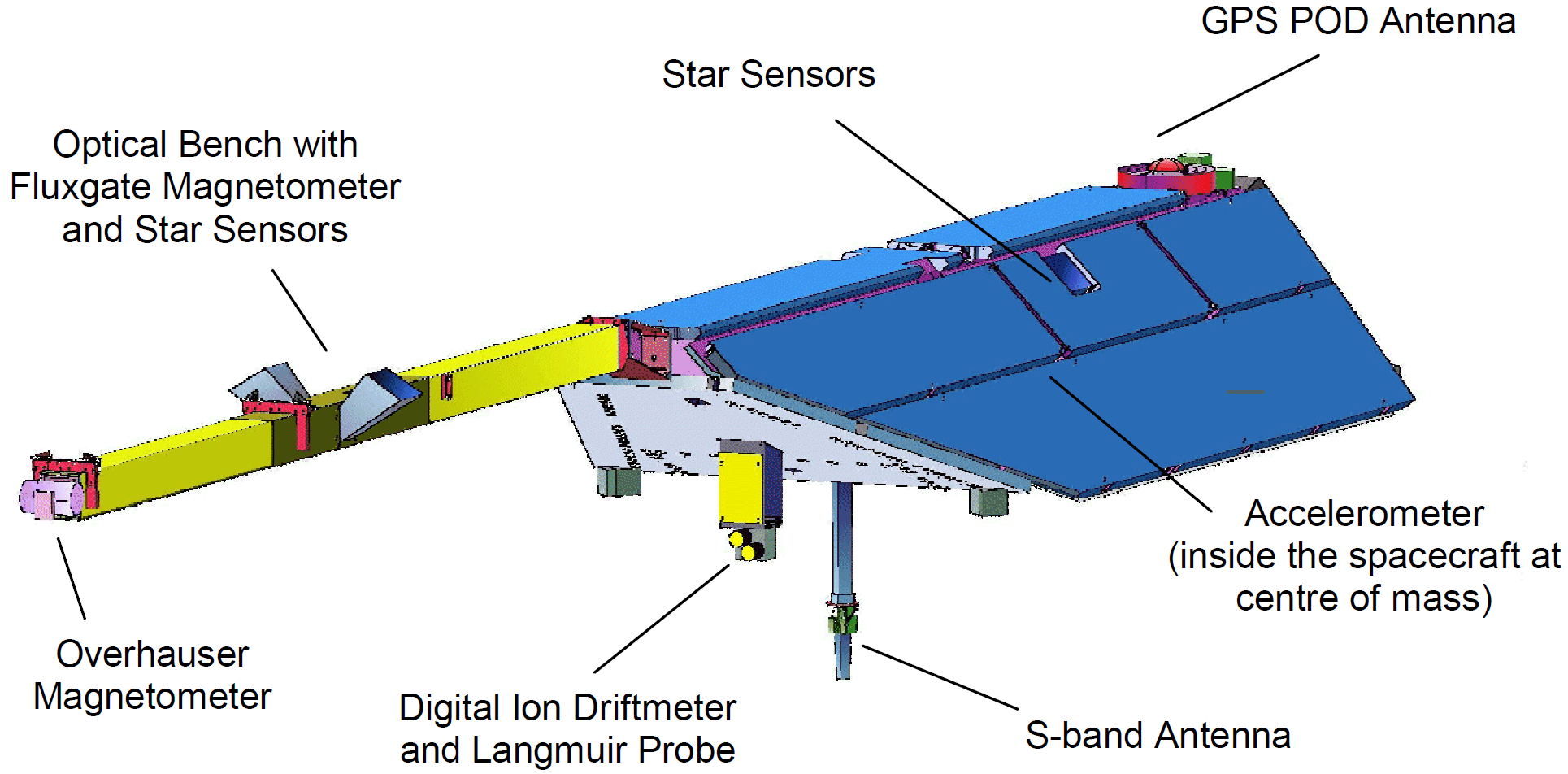}
\caption{Photograph of the CHAMP satellite~\cite{Reigber-2001-J}.}
\label{CHAMP}
\end{figure}

The CHAMP is a German small satellite mission for geoscientific and atmospheric research and applications, managed by GeoForschungsZentrum (GFZ)~\cite{Reigber-1999-J,Stolle-2006-J}. The satellite was built by the German space industry with the intent to foster high-tech capabilities, especially in the East-German space industry, and it was launched from the cosmodrome Plesetsk (north of Moscow) aboard a Russian COSMOS launch vehicle~\cite{Zhou-2018-J}. Fig.~\ref{CHAMP} shows a photograph of the CHAMP satellite. The payload includes an accelerometer, a magnetometer instrument assembly system, a GPS receiver, a laser retro reflector, and an ion drift meter. The magnetometer instrument assembly system is a boom-mounted package consisting of an Overhauser scalar magnetometer (the measurement range is 16000 nT $\sim$ 64000 nT, the resolution is 0.1 nT, the absolute accuracy is 0.5 nT, and the sampling rate is 1 Hz.)~\cite{Du-2016-J}, two fluxgate vector magnetometers (the measurement range is -64000 nT $\sim$ 64000 nT and the resolution is 1 nT $\sim$ 2 nT)~\cite{Korepanov-2012-J}, and two-star imagers to provide attitude information for fluxgates~\cite{Reigber-1996-C}.
\begin{figure}[htb]
\centering
\includegraphics[width=0.9\linewidth]{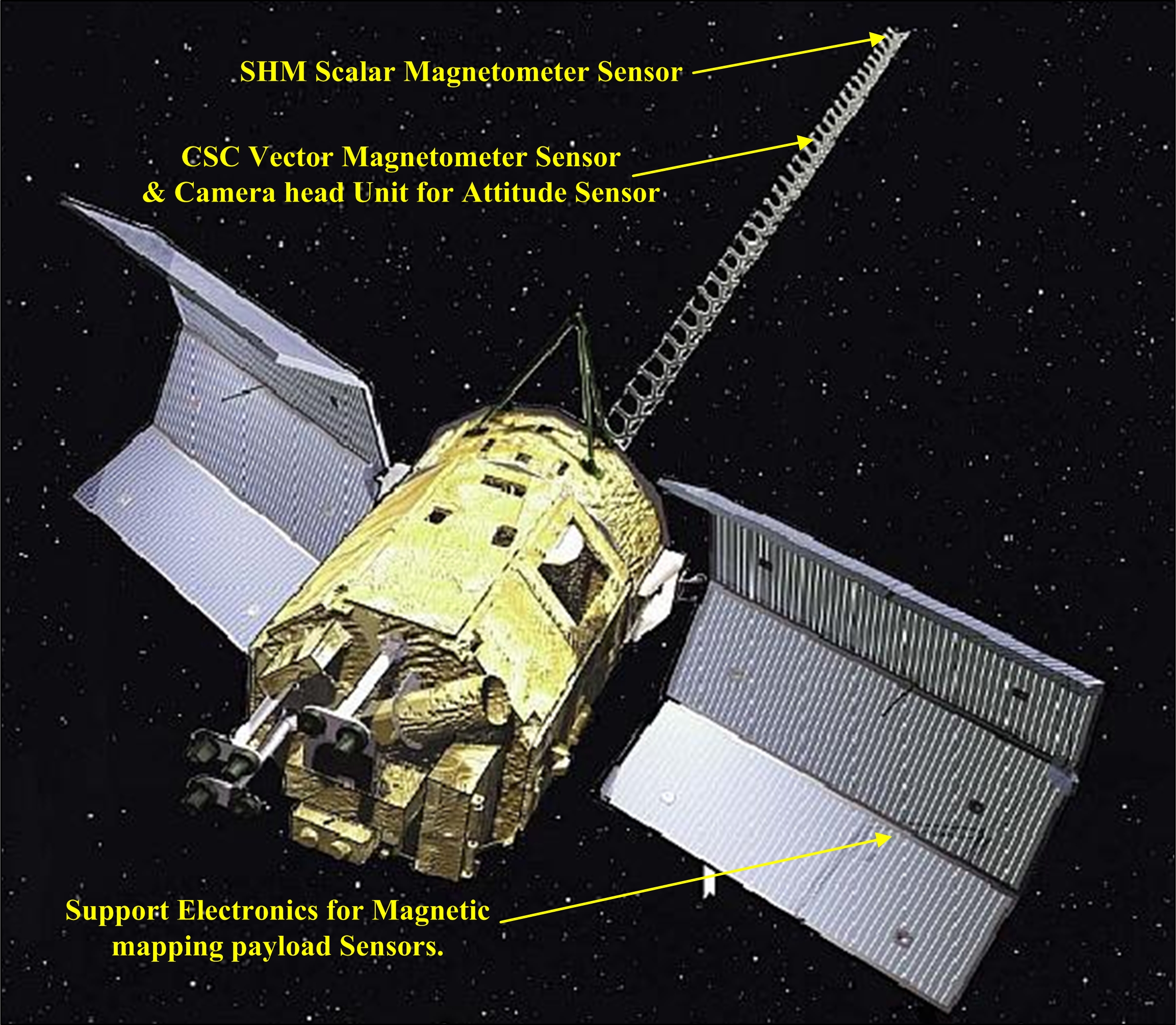}
\caption{Artist's view of the deployed SAC-C spacecraft~\cite{Colomb-2004-J}.}
\label{SAC-C}
\end{figure}

The SAC-C satellite (illustrated in Fig.~\ref{SAC-C}) is an international cooperative mission between the National Aeronautics and Space Administration (NASA), the Argentine Commission on Space Activities (CONAE), the French Space Agency, the Brazilian Space Agency, the Danish Space Research Institute, and the Italian Space Agency~\cite{Colomb-2002-J,Colomb-2003-J,Olsen-2006-J}. The SAC-C was developed through the partnership of its senior partners, CONAE and NASA with contributions from Brazil, Denmark, France, and Italy. It was the third satellite launched by CONAE and was the first operational Earth observation, designed to meet the requirements of socio-productive areas of our country, including agriculture, hydrology, coastlines, geology, health emergencies, etc. This satellite is equipped with a vector magnetometer and a star imager, further included are the associated support electronics to control the sensors and the boom deployment, a power control unit and a command and data handling unit. The geomagnetic measurement system on SAC-C represents essentially the same compact spherical coil based vector magnetometer as those flown on the ${\O}$rsted and CHAMP missions~\cite{Hajj-2004-J,Kuvshinov-2006-J,Alken-2007-J}. Besides, a scalar Helium magnetometer is placed on the tip of the boom and associated support electronics completes the magnetic mapping payload. The magnetic field measurements have a resolution of 1 nT at scalar and 2 nT at vector~\cite{Caruso-2000-J}.
\begin{figure*}[htb]
\centering
\includegraphics[width=1.0\linewidth]{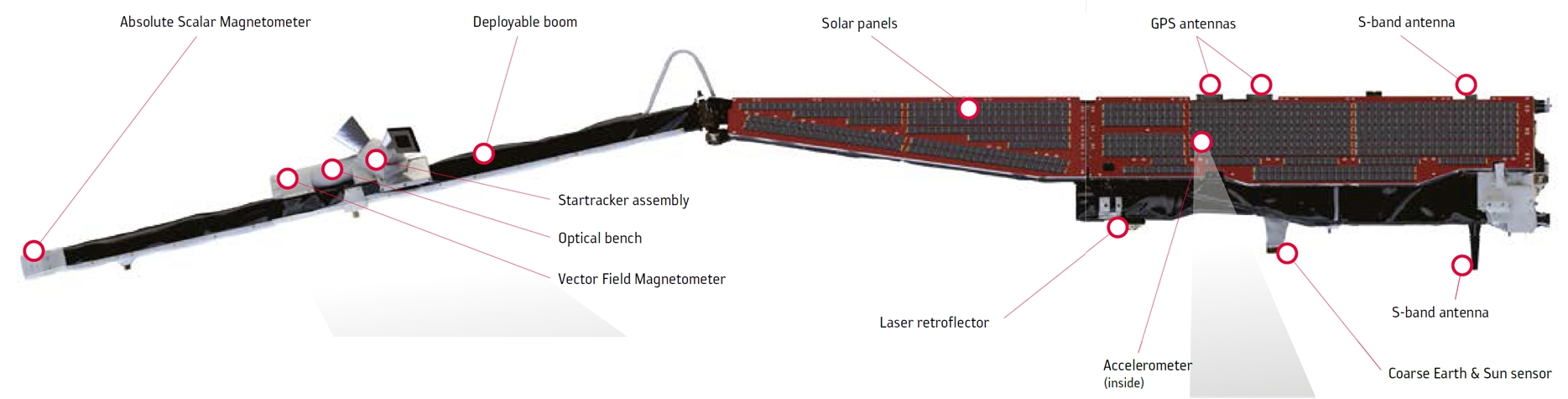}
\caption{Illustration of the Swarm satellite in orbit~\cite{Haagmans-2013-J}.}
\label{Swarm}
\end{figure*}

The Swarm satellite is constructed for the first constellation mission of the European Space Agency (ESA) for geomagnetic observation~\cite{Friis-2006-J}. This mission is operated by ESA's European Space Operations Centre (ESOC), in Germany, via the primary ground station in Kiruna, Sweden. The research objectives of the Swarm mission is to provide the best-ever survey of the geomagnetic field. Hence, a satellite constellation strategy is adopted. The satellite constellation is a group of artificial satellites working together as a system. Unlike a single satellite, a constellation can provide permanent global or near-global coverage, such that at any time everywhere on Earth at least one satellite is visible. Satellites are typically placed in sets of complementary orbital planes and connect to globally distributed ground stations. They may also use inter-satellite communication. This mission consists of the three identical Swarm satellites (A, B, and C), which were launched into a near-polar orbit~\cite{Friis-2008-J}. All the three Swarm satellites are equipped with the following set of identical instruments, including an absolute scalar magnetometer, a vector field magnetometer, a star tracker, an electric field instrument, a GPS receiver, a laser retro-reflector, and an accelerometer~\cite{Sabaka-2013-J, Xiong-2016-J}, as shown in Fig.~\ref{Swarm}. The absolute scalar magnetometer is an optically-pumped meta-stable helium-4 magnetometer, developed and manufactured by CEA-Leti under contract with the French Space Agency. It provides scalar measurements of the magnetic field to calibrate the vector field magnetometer. The vector field magnetometer is the mission's core instrument, which is developed and manufactured at the Technical University of Denmark~\cite{Jorgensen-2008-J}. It makes high-precision measurements of the magnitude and direction of the magnetic field. The orientation of the vector is determined by the star-tracker assembly, which provides attitude data. The vector field magnetometer and the star-trackers are both housed on an ultra-stable structure called an optical bench, halfway along the satellite's boom. 

\section{Discussion}
As the basic method of the vector magnetic survey technique, the ground vector magnetic survey development is the earliest and the technique is the most mature. The accuracy of the instrument system for the ground vector magnetic survey is high and the working performance is reliable. Nevertheless, there are also some problems, e.g., the detection depth is shallow and working efficiency is low. The wells vector magnetic survey is one useful tool for the Earth's deep mineral resources exploration. It can finish the detection work that the ground vector magnetic measurement can't do. However, due to the limit of the bore diameter and the high temperature in the borehole, the detection accuracy of the instrument system is lower than the ground magnetic instrument. The marine vector magnetic survey has an irreplaceable effect in the application of military or other ocean engineering. Its working mode changes from the method of ship drag-and-drop to small unmanned underwater vehicle, which can measure a more precise underwater magnetic anomaly. The airborne vector magnetic survey has the superiority of fast detection, high efficiency and strong practicality to the complex geophysical environment, etc. However, the detection accuracy is relatively low because of magnetic interference, the posture change of the aircraft carrier, etc. The study of magnetic compensation and data processing methods for aero three-component and full-tensor magnetic gradient is also needed. The satellites vector magnetic survey can obtain high-quality and whole global magnetic field data. It has the advantages of high detection efficiency, wide detection range, etc., accelerating and facilitating the investigations on space measurement technique and the evolution of the Earth's magnetic field.

In recent years, the multi-survey technique fusion approaches for geomagnetic surveys get a lot of attention, such as the aforementioned KTB drilling system. In the borehole, the fluxgate magnetometer is employed to collect the magnetic field data. For a better resolution of the magnetic anomaly, a detailed helicopter survey is carried out with the drill site in its center. In this case, complete coverage and high resolution of the magnetic field in the measuring level is obtained, which can obtain an excellent data set for analyzing the geological structure and supporting surface geological mapping. Further, there are some multi-survey data fusion based inversion methods, such as the joint inversion of surface and three-component borehole magnetic data~\cite{Li-2000-J}, the joint inversion of electromagnetic and magnetic data~\cite{Benech-2002-J}, etc., which can further improve the detection resolution for near-surface studies. However, there are still some problems with the multi-survey technique fusion that need to be further considered. To be specific, different magnetic vector survey techniques have to work simultaneously to achieve real-time synchronization of data measurement. In this case, these systems will be interfered by each other, causing the data abnormality. Furthermore, the problems of the consistency of data scale, the relevance ambiguity, and the data time alignment are all need to be further taken into account.  
 
\section{Conclusions and Expectations}
\label{sect:conclusion}
According to the investigations mentioned above, it can be realized that each magnetic survey method has its unique superiority and defect, and each single method has difficulty to meet the requirement of modern geophysical detection applications. With the development of the magnetic survey technique and instrument system, the future magnetic survey will change from the traditional work mode, i.e., measuring the total magnetic field intensity or its gradient to the vector information of the magnetic field, then to multi-parameter measurements. The magnetic survey application field will change from the traditional ground magnetic survey method with low efficiency and shallow detection depth to high efficient airborne magnetic survey, deep well magnetic survey, and abysmal sea magnetic survey, then gradually to the joint detection and interpretation combining with the five kinds of magnetic survey methods mentioned above. Through the combination of a variety of detection methods to realize advantageous complementarities, further improvement of detection accuracy and inversion resolution for Earth's magnetic field will be implemented.

Since the multi-survey technique fusion is immature, it suggests that fusion research should move from a feasibility study to problem-solving development. The research should demonstrate fusion as an effective solution in the context of a specific problem or application rather than simply showing a differently formulated fusion technique applicable to a number of general problems. To facilitate future research on multi-survey technique fusion at its initial stage, benchmark magnetic data and assessment protocols need to be created and established to fill the gap for evaluating different fusion methodologies. This remains a topic for our future work.

\section*{Acknowledgment}
This work is partly supported by the National Natural Science Foundation of China under Grant No. 41904164, the Natural Science Foundation of Hubei Province of China under Grant No. 2020CFB610, the Foundation of Wuhan Science and Technology Bureau under Grant No. 2019010701011411, the Foundation of National Key Research and Development Program of China under Grant No. 2018YFC1503702, and the Fundamental Research Funds for the Central Universities, China University of Geosciences (Wuhan) under Grant No. CUG190628.

\bibliographystyle{IEEEtranTIE}
\bibliography{BIB_1x-TIE-2xxx}

\end{document}